\documentclass[12pt,onecolumn,article]{IEEEtran}
\usepackage{cite}

\usepackage[cmex10]{amsmath}
\usepackage{enumerate}
\usepackage{tikz}
\usepackage{amsmath}
\usepackage{amssymb}
\usepackage{amsthm}
\usepackage{mathrsfs}

\RequirePackage{mleftright} %provides \mleft, \mright to fix issue with spacing
\mleftright                 %redefines all \left and \right to behave
\newcommand{\markov}{\textnormal{\mbox{$\multimap\hspace{-0.73ex}-\hspace{-2ex}-$}}}

\newcommand{\Rc}{R_\textnormal{c}}
\newcommand{\Rpr}{R_\textnormal{p}}
\newcommand{\CSh}{C_{\textnormal{Sh}}}
\newcommand{\Cdet}{{\mathscr C}_{\textnormal{det}}}
\newcommand{\CdetI}{{\mathscr C}_{\textnormal{det}}^{(I)}}
\newcommand{\CdetO}{{\mathscr C}_{\textnormal{det}}^{(O)}}

\newcommand{\MMc}{\set{M}_{\textnormal{c}}}
\newcommand{\MMpr}{\set{M}_{\textnormal{p}}}
\newcommand{\Mc}{M_{\textnormal{c}}}
\newcommand{\Mpr}{M_{\textnormal{p}}}
\newcommand{\mc}{m_{\textnormal{c}}}
\newcommand{\mpr}{m_{\textnormal{p}}}
\newcommand{\mchat}{\hat{m}_{\textnormal{c}}}
\newcommand{\mctilde}{\tilde{m}_{\textnormal{c}}}
\newcommand{\mprhat}{\hat{m}_{\textnormal{p}}}
\newcommand{\phiz}{\phi_{z}}
\newcommand{\phiy}{\phi_{y}}

\newcommand{\Hbin}[1]{H_{\text{b}} (#1)}
\newcommand{\pmf}[1]{\mat{#1}}% For PMFs.
\newcommand{\ch}[1]{\mat{#1}} % For channels
\newcommand{\BSC}[1]{\textnormal{BSC}(#1)}

\newcommand{\bfy}{\mathbf y}

\newcommand{\pmin}{p_{\textnormal{min}}}
\newcommand{\pmax}{p_{\textnormal{max}}}

\newcommand{\bfx}{{\mathbf x}}

\newcommand{\bfs}{{\mathbf s}}

\newcommand{\Code}{{\mathcal C}}

\newcommand{\eps}{\epsilon}
\newcommand{\bfz}{{\mathbf z}}
\newcommand{\abs}[1]{\lvert#1\rvert} %absolute value 
\newcommand{\card}[1]{\abs{#1}} %Cardinality of set
\newcommand{\set}[1]{\mathcal{#1}} % Sets in calligraphic font
\newcommand{\mat}[1]{\mathsf{#1}}
\newtheorem{theorem}{Theorem}%[section]
\newtheorem{proposition}[theorem]{Proposition}

\newtheorem{remark}[theorem]{Remark}

\begin{document}
\title{Semi-Robust Communications over a Broadcast Channel}
\author{Tibor Keresztfalvi and Amos Lapidoth}
\maketitle
\begin{abstract}
  We establish the deterministic-code capacity region of a network
  with one transmitter and two receivers: an ``ordinary receiver'' and
  a ``robust receiver.'' The channel to the ordinary receiver is a
  given (known) discrete memoryless channel, whereas the channel to
  the robust receiver is an arbitrarily varying channel. Both
  receivers are required to decode the ``common message,'' whereas
  only the ordinary receiver is required to decode the ``private
  message.''
\end{abstract}
\begin{IEEEkeywords}
  Arbitrarily varying channel,
  % AVC,
  binary symmetric,
  broadcast channel,
  %common message,
  degraded message set,
  % private message,
  robust communications.   
\end{IEEEkeywords}
\section{Introduction}

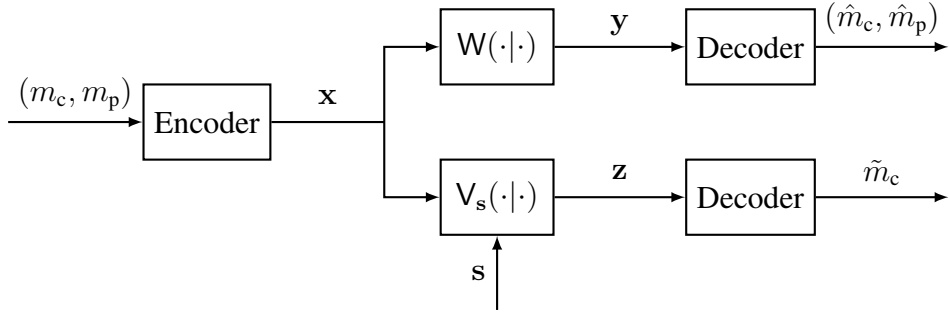
\begin{figure}
\makebox[\textwidth][c]{
\begin{tikzpicture}[scale=1,>=latex,thick]

%Styles
\tikzstyle{mybox1}=[draw, rectangle,  minimum height=1cm, minimum
width=1.5cm]
\tikzstyle{mybox2}=[draw, rectangle,  minimum height=3cm,anchor=north west,text width=1.5cm]

% Nodes
\node[mybox1, anchor=east] at (0,0) (enc) {Encoder};
\node[mybox1, anchor=center] at (3,1) (dmc) {$\ch{W} (\cdot | \cdot)$};
\node[mybox1, anchor=center] at (3,-1) (avc) {$\ch{V}_{\bfs} (\cdot |
  \cdot)$};
\node[mybox1, anchor=west] at (5.5,1) (decdmc) {Decoder};
\node[mybox1, anchor=west] at (5.5,-1) (decavc) {Decoder};
\coordinate (branch) at (1.5,0);
\coordinate (state) at (3,-2.5);

% Connections
\draw[->] (-3.5,0) -- node[above] {$(\mc, \mpr)$} (enc);
\draw (enc) -- node[above] {$\phantom{\bfy} \bfx \phantom{\bfy}$} (branch);
\draw[->] (branch) |- (dmc);
\draw[->] (branch) |- (avc);
\draw[->] (dmc) -- node[above] {$\bfy$} (decdmc);
\draw[->] (avc) -- node[above] {$\phantom{\bfy} \bfz \phantom{\bfy}$} (decavc);
\draw[->] (decdmc) -- node[above] {$(\mchat, \mprhat)$} (9,1);
\draw[->] (decavc) -- node[above] {$\mctilde$} (9,-1);
\draw[->] (state) -- node[left] {$\bfs$} (avc);

\end{tikzpicture}}
\caption{The semi-arbitrarily-varying broadcast channel (semi-AVBC)
  with common message $\mc$, private message $\mpr$, and state
  sequence $\bfs\in\set{S}^n$.
  % , and a blocklength-$n$ encoder.
%  The decoders produce the message estimates $(\mchat, \mprhat)$ and
%  $\mctilde$ respectively.
  \label{fig:savbc}
}
\end{figure}

As in Figure~\ref{fig:savbc}, two independent data streams---a
rate-$\Rc$ common data stream and a rate-$\Rpr$ private data
stream---%
% the of rates $\Rc$ and $\Rpr$
are to be transmitted over a broadcast channel with two receivers: an
``ordinary receiver'' and a ``robust receiver.'' The channel to the
ordinary receiver, the receiver that is required to recover both
streams reliably, is a given (known) discrete memoryless channel (DMC)
$\ch{W} (y|x)$. The channel to the robust receiver, the receiver that
is only required to recover the common stream, is an arbitrarily
varying channel (AVC) \cite{Lapidoth_Prakash98}. The set of rate pairs
$(\Rc,\Rpr)$ that can be communicated reliably under these
requirements is the \emph{capacity region}, which we derive here.

%The ordinary receiver is required
%to reliably recover both data streams, whereas the robsust receiver is
%only required to recover and the receiver with unknown channel
%statistics is required to reconstruct the first data stream.

The scenario where one receiver must recover both streams and the
other only one, falls under the heading of \emph{degraded message sets.}
The capacity region of the broadcast channel with degraded message
sets was established by K\"orner and Marton
in~\cite{KornerMarton77}. But their model differs from ours because
their broadcast channel is fixed and given: there is nothing ``varying''
about it.

Our network can be viewed as an arbitrarily varying broadcast channel
(AVBC) of a special kind: one where the channel to one of the
receivers is degenerate in the sense of being given and not depending
on the state. General AVBCs where studied by Jahn~\cite{Jahn81} who
derived an inner bound on their capacity regions, and our
achievability result essentially follows from his.
% (with a slight random time-sharing twist).
Our converse shows that in our setting the inner bound is tight.

More recent results on the AVBC for settings with causal and noncausal
side information were obtained by 
% was first
% considered in~\cite{Blackwell60}. An inner bound on the random code
% capacity of the AVBC without side information was given by Jahn
% in~\cite{Jahn81}. In recent works by
Pereg and Steinberg~\cite{PeregAVBCDMS,PeregDAVBC,Steinberg05}.

%We establish the random code capacity region of the partially
%arbitrarily varying broadcast channel (SAVBC) with degraded message
%sets.

\section{The Main Result}
A discrete memoryless \emph{state-dependent broadcast channel}
$(\set{X}, \set{Y}, \set{Z}, \set{S}, \ch{W}_{Y,Z|X,S})$ consists of a
finite input alphabet $\set{X}$, finite output alphabets $\set{Y}$ and
$\set{Z}$, a (not necessarily finite) state set $\set{S}$, and a
collection of transition probability matrices $\ch{W}_{Y,Z|X,S}$. A
\emph{semi-AVBC} (SAVBC) is a state-dependent broadcast channel where
the conditional law of the output $Y$ given the input $x$ and the
state $s$ does not depend on the state. For such a channel, we denote the marginal
conditional distributions of the outputs $Y$ and $Z$ given the input
$x$ and the state $s$ by $\ch{W} (y|x)$ and $\ch{V}_s (z|x)$ respectively:
\begin{subequations}
\begin{IEEEeqnarray}{rCl}
  \ch{W} (y|x)  & = & \ch{W}_{Y|X,S} (y|x,s), \\
  \ch{V}_s (z|x) & = & \ch{W}_{Z|X,S} (z|x,s).
\end{IEEEeqnarray}
\end{subequations}
Given a blocklength $n$, an input
sequence $\bfx \in \set{X}^{n}$, and a state sequence $\bfs \in \set{S}^{n}$,
\begin{IEEEeqnarray}{rCl}
  \ch{W}_{Y^{n},Z^n|X^n,S^n} (\bfy, \bfz|\bfx, \bfs) = \prod_{i=1}^n
  \ch{W}_{Y,Z|X,S} (y_i,z_i|x_i,s_i), \quad (\bfy,\bfz) \in
  \set{Y}^{n} \times \set{Z}^{n}.
\end{IEEEeqnarray}

We consider the transmission from \emph{degraded message sets}: the
encoder sends a \emph{common message} $\mc$ to both receivers and a
\emph{private message} $\mpr$ to the receiver observing~$Y$. The receiver observing $Z$ is
thus only required to decode the common message.

Given a blocklength $n$, a \emph{deterministic code} $\Code$ for the
SAVBC consists of a common message set $\MMc$ with $2^{n\Rc}$
messages, a private message set $\MMpr$ with $2^{n\Rpr}$ messages,
an encoder mapping
\begin{IEEEeqnarray}{rCl}
  f\colon\, \MMc\times\MMpr \to \set{X}^n,
\end{IEEEeqnarray}
and decoding mappings
\begin{subequations}
  \label{block:encoders}
\begin{IEEEeqnarray}{rCl}
  \phiy\colon\, \set{Y}^n &\to& \MMc\times\MMpr \\
  \phiz\colon\, \set{Z}^n &\to& \MMc.
\end{IEEEeqnarray}
\end{subequations}
%The common-message rate is denoted $\Rc$, and the private-message rate
%$\Rpr$.
The message-averaged probability of error of a code $\Code$ given a
state sequence $\bfs \in \set{S}^n$ is
\begin{IEEEeqnarray}{rCl}
  P_{\textnormal{e}|\bfs}^{(n)} (\Code) =
  \frac{1}{\card{\MMc}\card{\MMpr}}
  \sum_{(\mc,\mpr)\in\MMc\times\MMpr} \sum_{(\bfy,\bfz) \not\in
    \set{D}(\mc, \mpr)} \ch{W}_{Y^n,Z^n|X^n,S^n} (\bfy, \bfz|\bfx,
  \bfs),
\end{IEEEeqnarray}
where
\begin{IEEEeqnarray}{rCl}
  \set{D} (\mc, \mpr) = \bigl\{ (\bfy, \bfz) \in \set{Y}^n\times\set{Z}^n\,
  \colon \phiy (\bfy) = (\mc, \mpr),\, \phiz (\bfz) = \mc \bigr\}.
\end{IEEEeqnarray}
% The average probability of error of a code $\Code$ corresponding to a
% distribution $p_{S^n}$ on $\set{S}^n$ is then defined as
% \begin{IEEEeqnarray}{rCl}
%   P_{\textnormal{e}}^{(n)} (p_{S^n}, \Code) = \sum_{\bfs\in\set{S}^n}
%   p_{S^n} (\bfs)\, P_{\textnormal{e}|\bfs}^{(n)} (\Code).
% \end{IEEEeqnarray}
We say that the rate pair $(\Rc, \Rpr)$ is \emph{achievable with
deterministic codes}, if there exists a sequence of codes $\{\Code_{n}\}$ with
rates $(\Rc, \Rpr)$ such that
\begin{IEEEeqnarray}{rCl}
  \lim_{n \to \infty} \; \sup_{\bfs \in \set{S}^{n}}
  P_{\textnormal{e}|\bfs}^{(n)} (\Code_{n}) = 0.
  %\le \eps_{n},
%  P_{\textnormal{e}}^{(n)} (p_{S^n}, \Code) \le \eps_n,
\end{IEEEeqnarray}
% for all PMFs $p_{S^n}$ on $\set{S}^n$, and
% where $\eps_n$ tends to zero as $n$ tends to infinity.
The \emph{deterministic code capacity} $\Cdet$ (under the
average-probability-of-error criterion) of the SAVBC is
% defined as
the closure of the set of rate pairs that are achievable with
deterministic codes.

As in~\cite[Corollary~12.3]{CK11}, it can be shown that the capacity
region depends on the states only via the convex-closure of the
channels they induce. We shall thus make the following assumption
without any loss of generality:
\medskip

\noindent\textbf{Assumption}: We shall assume throughout that
$\{\ch{V}_{s}(z|x)\}_{s \in \set{S}}$ is compact\footnote{If the set
  $\{\ch{V}_{s}(z|x)\}_{s \in \set{S}}$ is not compact our result
  still holds, but with infima replacing the minima in the
  characterizations of the capacity region.} and convex
%\footnote{A similar achievable region can be shown if we allow
%  any set $\set{S}$, and define $\bar{\set{S}}$ as the set of all
%  convex combinations of $s\in\set{S}$ in the above mentioned sense.}
in the sense that for every $0 < \lambda < 1$ and
$s_1, s_2 \in \set{S}$, there exists a state $\bar{s}\in\set{S}$ such
that
\begin{IEEEeqnarray}{rCl}
  \ch{V}_{\bar{s}}(z|x) = \lambda \, \ch{V}_{s_1}(z|x) + (1-\lambda)
  \, \ch{V}_{s_2}(z|x), \quad (x,z) \in \set{X} \times \set{Z}.
\end{IEEEeqnarray}
\medskip

Following \cite[Remark~IIB2]{Jahn81} or using a time-sharing argument
we note:

\medskip
  %allows us to determine whether the interior of $\Cdet$ is
  %nonempty:
  \begin{remark}\label{rem:nonemptyint}
    The interior of $\Cdet$ is nonempty if, and only if, the capacity
    of the channel $\ch{W}(y|x)$ to $Y$ and the capacity (under the
    average-probability-of-error criterion) of the AVC to $Z$ are both
    positive. The latter is positive if, and only if, the AVC is
    \emph{nonsymmetrizable} \cite{CN88,Csi92}.
  \end{remark}

  \medskip

  We next define the region $\CdetI$ that will turn out to equal the
  capacity region when the latter is not empty. It is defined as the
  closure of the union over all PMFs $p_{U,X}$ of the set of rate
  pairs $(\Rc, \Rpr)$ that satisfy
  \begin{subequations}
  \label{block:cap_noQ}
  \begin{IEEEeqnarray}{rCl}
    \Rc &\le& \min_{s\in\set{S}} I(U;Z) \label{eq:noQ_amos104} \\
    \Rpr &\le& I(X;Y|U) \label{eq:noQ_amos105} \\
    \Rc + \Rpr &\le& I(X;Y),
  \end{IEEEeqnarray}
\end{subequations}
where the mutual informations are computed w.r.t.\ the joint
  distribution
  \begin{IEEEeqnarray}{rCl}
    p_{U,X} (u,x)\, \ch{W} (y|x)\, \ch{V}_s (z|x),
    \label{eq:noQ_factorizationForm}
  \end{IEEEeqnarray}
  and where $U$ is an auxiliary chance variable taking values in a
  finite set $\set{U}$. Our main result is the following theorem.

  \medskip

  \begin{theorem}\label{thm:rcapreg}
    Under the above assumption, if the deterministic-code capacity
    $\Cdet$ of a SAVBC is not empty, then it equals $\CdetI$:
  \begin{IEEEeqnarray}{rCl}
    \Bigl( \Cdet \neq \emptyset \Bigr) \implies \Bigl( \Cdet = \CdetI \Bigr).
  \end{IEEEeqnarray}
\end{theorem}

\medskip

%   \begin{theorem}\label{thm:rcapreg}
%   The random code capacity $\Crnd$ of a SAVBC with degraded message
%   sets is equal to $\CrndI$:
%   \begin{IEEEeqnarray}{rCl}
%     \Crnd & = & \CrndI
%   \end{IEEEeqnarray}
% \end{theorem}

\section{Proof of the Main Result}
The achievability result---that $\Cdet \neq \emptyset$ implies that
every rate pair $(\Rc, \Rpr)$ satisfying~\eqref{block:cap_noQ} for
some $p_{U,X}$ is achievable---follows directly from
Jahn~\cite[Theorem~2]{Jahn81}. We therefore focus on the converse,
i.e., on showing that the achievability of a rate pair $(\Rc, \Rpr)$
implies that it lies in $\CdetI$. But before proving this, we study
$\CdetI$.

\medskip

\begin{proposition}
  \label{prop:withQ}The region~$\CdetI$ can also be expressed as the
  closure of the union over all PMFs $p_{U,X,Q}$ of the set of rate pairs
  $(\Rc, \Rpr)$ that satisfy
  \begin{subequations}
  \label{block:capreg}
  \begin{IEEEeqnarray}{rCl}
    \Rc &\le& \min_{s\in\set{S}} I(U;Z|Q) \label{eq:amos104} \\
    \Rpr &\le& I(X;Y|U,Q) \label{eq:amos105} \\
    \Rc + \Rpr &\le& I(X;Y|Q),
  \end{IEEEeqnarray}
\end{subequations}
where the mutual informations are computed w.r.t.\ the joint
  distribution
  \begin{IEEEeqnarray}{rCl}
    p_{U,X,Q} (u,x,q)\, \ch{W} (y|x)\, \ch{V}_s (z|x),
    \label{eq:factorizationForm}
  \end{IEEEeqnarray}
  and where $U$ and $Q$ are auxiliary chance variables taking values
  in the finite sets $\set{U}$ and $\set{Q}$.
  \end{proposition}

\medskip

\begin{IEEEproof}
    One inclusion is obvious and simply follows by setting $Q$ to be
    deterministic. We therefore focus on the other, namely, on showing
    that if there exists some joint PMF $p_{U,X,Q}$ under which the
    pair $(\Rc, \Rpr)$ satisfies~\eqref{block:capreg}, then there exists
    some auxiliary chance variable $\tilde{U}$ and a PMF
    $p_{\tilde{U},X}$ under which the pair
    satisfies~\eqref{block:cap_noQ} when we substitute $\tilde{U}$ for
    $U$. To this end we choose $\tilde{U} = (U,Q)$ and show that the
    results of substituting $\tilde{U}$ for $U$ on the RHS of each of
    the inequalities in~\eqref{block:cap_noQ} is at least as high as
    the RHS of the corresponding inequality in~\eqref{block:capreg}:
    \begin{IEEEeqnarray*}{rCl}
      \min_{s\in\set{S}} I(\tilde{U};Z) & = &  \min_{s\in\set{S}}
      I(U,Q;Z) \\
      & = & \min_{s\in\set{S}} \bigl\{ I(U;Z|Q) + I(Q;Z) \bigr\}\\
      & \geq & \min_{s\in\set{S}} I(U;Z|Q);
    \end{IEEEeqnarray*}
    \begin{IEEEeqnarray*}{rCl}
      I(X;Y|\tilde{U}) &= & I(X;Y|U,Q);
    \end{IEEEeqnarray*}
    and
    \begin{IEEEeqnarray}{rCl}
      I(X;Y) & = & I(X,Q;Y) \label{eq:stam10}\\
      & = & I(Q;Y) + I(X;Y|Q) \nonumber \\
      & \ge & I(X;Y|Q), \nonumber
    \end{IEEEeqnarray}
    where %the first equality
    \eqref{eq:stam10} follows from the Markovity
    $Q \markov X \markov Y$.
  \end{IEEEproof}
  From Proposition~\ref{prop:withQ} we obtain:

  \medskip
  
  \begin{proposition}\label{prop:convex}
    The region $\CdetI$ %of Theorem~\ref{thm:rcapreg}
    is a compact convex set containing the rate pairs
  \begin{subequations}
    \begin{IEEEeqnarray}{rCl}
      \label{eq:bath100}
    \Big( \min \bigl\{ \CSh (\ch{W}), \min_{s\in\set{S}} \CSh (\ch{V}_s)
    \bigr\},\, 0 \Big)
  \end{IEEEeqnarray}
  and
  \begin{IEEEeqnarray}{rCl}
    \label{eq:bath110}
    \big( 0,\, \CSh (\ch{W}) \big),
  \end{IEEEeqnarray}
\end{subequations}
  where $\CSh (\ch{W})$ denotes the Shannon capacity of the channel
  $\ch{W}$. Moreover, $\CdetI$ is included in the triangle with
  vertices
  \begin{IEEEeqnarray}{rCl}
    \label{eq:convtri}
    (0, 0), \quad
    (\CSh (\ch{W}), 0), \quad
    (0, \CSh (\ch{W})).
  \end{IEEEeqnarray}
\end{proposition}
\begin{IEEEproof}
  The convexity is due to the auxiliary chance variable $Q$. To see
  that the pair~\eqref{eq:bath100} is in $\CdetI$, consider choosing
  $Q$ to be deterministic and $U$ to equal $X$. To see that the
  pair~\eqref{eq:bath110} is in $\CdetI$, consider choosing both $U$
  and $Q$ to be deterministic. The inclusion in the
  triangle~\eqref{eq:convtri} follows from
  $I(X;Y|Q) \le \CSh (\ch{W})$.
  \end{IEEEproof}

  As a final step before proving the converse, we next provide one
  last characterization of $\CdetI$. To that end, let~$\CdetO$ denote
  the set of rate pairs $(\Rc, \Rpr)$ that satisfy
\begin{subequations}
\label{block:eqvrregion}
\begin{IEEEeqnarray}{rCl}
  \Rc & \le & \min_{s\in\set{S}} I(U;Z|Q) \label{eq:amos107a}\\
  \Rpr + \Rc & \le & I(X;Y|U,Q) + \min_{s\in\set{S}}
  I(U;Z|Q) \label{eq:amos107} \\
  \Rpr + \Rc & \le & I(X;Y|Q), \label{eq:amos107c}
\end{IEEEeqnarray}
for some PMF $p_{U,X,Q}$, where the mutual informations are computed
w.r.t.\ the joint PMF of~\eqref{eq:factorizationForm}.
% \begin{IEEEeqnarray}{rCl}
%    p_{U,X,Q}(u,x,q)\, \ch{W}(y|x) \ch{V}_{s}(z|x).
% \end{IEEEeqnarray}
\end{subequations}
The inequalities in the definition of $\CdetO$ thus differ from those
in \eqref{block:capreg} in that we have replaced~\eqref{eq:amos105}
with~\eqref{eq:amos107}.  As we shall next show, this replacement does
not change the region, and $\CdetO = \CdetI$. Once we show this, we
will prove the converse for~$\CdetO$.
%Instead of proving the converse for $\Crnd$, we prove it for the
%equivalent region $\set{R}$.

\begin{proposition}
  \label{prop:R_is_CdetI}
  The region $\CdetO$, which is defined in~\eqref{block:eqvrregion},
  is equal to $\CdetI$:
  %, which is defined in~\eqref{block:capreg} are equal
  \begin{equation}
    \CdetO = \CdetI.
  \end{equation}
\end{proposition}
\begin{IEEEproof}
  We shall prove this result using the characterization of $\CdetI$ of
  Proposition~\ref{prop:withQ}.  To see why the two regions are
  equivalent, fix some PMF $p_{U,X,Q}$ and consider the
  bounds in~\eqref{block:capreg} and~\eqref{block:eqvrregion}. If a rate
  pair satisfies~\eqref{block:capreg}, then it also
  satisfies~\eqref{block:eqvrregion}, because~\eqref{eq:amos107} is
  the result of adding~\eqref{eq:amos104}
  and~\eqref{eq:amos105}. Thus, $\CdetI \subseteq \CdetO$.
% \begin{IEEEeqnarray}{rCl}
%   \label{eq:incl1}
%   \CrndI \subseteq \set{R}.
% \end{IEEEeqnarray}
  
To establish the reverse inclusion we consider two cases separately.

\noindent
\emph{Case I: $\min_{s\in\set{S}} I(U;Z|Q) + I(X;Y|U,Q) \geq I(X;Y|Q)$.}

\noindent In this case Inequality~\eqref{eq:amos107} is implied
by~\eqref{eq:amos107c} and is thus redundant. Consequently, we
need to show that the trapezoid defined by~\eqref{eq:amos107a}
and~\eqref{eq:amos107c} of vertices
\begin{equation*}
  (0,0), \quad
  \bigl(\min_{s\in\set{S}} I(U;Z|Q),0\bigr), \quad
  \bigl(\min_{s\in\set{S}} I(U;Z|Q), \; I(X;Y|Q) - \min_{s\in\set{S}}
  I(U;Z|Q) \bigr),
  \quad
  \bigl(0, I(X;Y|Q)\bigr)
\end{equation*}
is contained in $\CdetI$. This can be shown by noting that $\CdetI$
contains the pentagon defined by \eqref{block:capreg} of vertices
\begin{multline*}
    (0,0), \quad
  \bigl(\min_{s\in\set{S}} I(U;Z|Q),0\bigr), \quad
  \bigl(\min_{s\in\set{S}} I(U;Z|Q), \; I(X;Y|Q) - \min_{s\in\set{S}}   I(U;Z|Q) \bigr),
  \\
  \bigl(I(X;Y|Q) - I(X;Y|U,Q), \; I(X;Y|U,Q)\bigr), \quad
  \bigl( 0, I(X;Y|U,Q) \bigr);
\end{multline*}
that $\CdetI$ also contains the point $\{0, I(X;Y|Q)\}$
(Proposition~\ref{prop:convex}); and that it thus also contains the
convex hull of the union of the pentagon and the point
(Proposition~\ref{prop:convex}).

\noindent
\emph{Case II: $\min_{s\in\set{S}} I(U;Z|Q) + I(X;Y|U,Q) <  I(X;Y|Q)$.}

\noindent In this case~\eqref{eq:amos107c} is implied
by~\eqref{eq:amos107}, and we need to show that the trapezoid defined
by~\eqref{eq:amos107a} and~\eqref{eq:amos107} of vertices
\begin{equation*}
  (0,0), \quad
  \bigl(\min_{s\in\set{S}} I(U;Z|Q),0\bigr), \quad
  \bigl(\min_{s\in\set{S}} I(U;Z|Q), I(X;Y|U,Q)\bigr), \quad
  \bigl(0, \; I(X;Y|U,Q) + \min_{s\in\set{S}} I(U;Z|Q) \bigr),
\end{equation*}
is contained in $\CdetI$. This can be shown by noting that $\CdetI$
contains the rectangle defined by \eqref{block:capreg} of vertices
\begin{equation*}
  (0,0), \quad
  \bigl(\min_{s\in\set{S}} I(U;Z|Q),0\bigr), \quad
  \bigl(\min_{s\in\set{S}} I(U;Z|Q), I(X;Y|U,Q)\bigr), \quad
  \bigl(0, \; I(X;Y|U,Q) \bigr);
\end{equation*}
it contains the point $\{(0,I(X;Y|U,Q) + \min_{s\in\set{S}}
I(U;Z|Q))\}$ (because in the case under consideration  $I(X;Y|U,Q) +
\min_{s\in\set{S}} I(U;Z|Q)$ is smaller than $I(X;Y|Q)$ so the
achievability of this point follows from Proposition~\ref{prop:convex}); and
because $\CdetI$ is convex.
\end{IEEEproof}

% On the other hand, it follows from Remark~\ref{rem:convex} that if
% \begin{IEEEeqnarray}{rCl}
%   \min_{s\in\set{S}} I(U;Z|Q) + I(X;Y|U,Q) > I(X;Y|Q),
% \end{IEEEeqnarray}
% then the time-sharing region between the rate pairs
% \begin{IEEEeqnarray}{rCl}
%   \big( 0, \CSh (\ch{W}) \big) \nonumber
% \end{IEEEeqnarray}
% and
% \begin{IEEEeqnarray}{rCl}
%   \big( \min_{s\in\set{S}} I(U;Z|Q),\, \min_{s\in\set{S}} I(U;Z|Q) +
%   I(X;Y|U,Q) - I(X;Y|Q) \big) \nonumber
% \end{IEEEeqnarray}
% is included in $\Crnd$ for every $p_{U,X,Q}$, and if
% \begin{IEEEeqnarray}{rCl}
%   \min_{s\in\set{S}} I(U;Z|Q) + I(X;Y|U,Q) \le I(X;Y|Q),
% \end{IEEEeqnarray}
% then the time-sharing region between the rate pairs
% \begin{IEEEeqnarray}{rCl}
%   \big( 0, \CSh (\ch{W}) \big) \nonumber
% \end{IEEEeqnarray}
% and
% \begin{IEEEeqnarray}{rCl}
%   \big( \min_{s\in\set{S}} I(U;Z|Q),\, I(X;Y|U,Q) \big) \nonumber
% \end{IEEEeqnarray}
% is included in $\Crnd$ for every $p_{U,X,Q}$. Consequently, if any
% rate pair satisfies~\eqref{block:eqvrregion} for a PMF $p_{U,X,Q}$, it
% is included in $\set{R}$ and hence
% \begin{IEEEeqnarray}{rCl}
%   \label{eq:incl2}
%   \Crnd \supseteq \set{R}.
% \end{IEEEeqnarray}
% It follows from~\eqref{eq:incl1} and \eqref{eq:incl2}, that the two
% regions are equivalent.

We shall now prove the converse by proving that no rate pair outside
$\CdetO$ is achievable.

\medskip

\begin{IEEEproof}[Proof of the converse part of {Theorem~\ref{thm:rcapreg}}]
  We first note that, as in \cite[Theorem~15.6.1]{CT2}, the capacity
  region depends only on the marginals (namely the given channel
  $\ch{W}(y|x)$ and the AVC). There is thus no loss in generality in
  assuming, as we shall, that $Y$ and $Z$ are conditionally
  independent given the channel input and state.

Fix finite sets $\MMc$, $\MMpr$, a blocklength-$n$ $(n, \MMc, \MMpr)$ encoder
\begin{IEEEeqnarray}{rCl}
  f \colon \MMc \times \MMpr \to \set{X}^n
\end{IEEEeqnarray}
and decoders as in~\eqref{block:encoders} with
$\sup_{\bfs \in \set{S}^{n}} P_{\textnormal{e}|\bfs}^{(n)} (\Code)$
tending to zero. Fix also a state sequence
$\bfs = (s_1,\ldots , s_n)$.
%which is possibly the outcome of a randomized encoder.
%
%\texttt{We might have to make the converse for deterministic encoders only}
%
%that is chosen with
%positive probability from the collection of codes from which the
%encoder and decoders choose randomly, with encoder mapping
Draw the message pair $(\Mc, \Mpr)$ uniformly over $\MMc\times\MMpr$,
and denote its distribution $p_{\Mc,\Mpr}$.
%Let $\bfs = (s_1,\ldots , s_n)$ denote the state sequence of the BC. Then,
In view of our conditional independence assumption, for every
$(\mc, \mpr, \bfx, \bfy, \bfz) \in \MMc\times\MMpr\times\set{X}^n
\times \set{Y}^n \times \set{Z}^n$
\begin{IEEEeqnarray}{rCl}
  \IEEEeqnarraymulticol{3}{l}{\pmf{P} [(\Mc, \Mpr, X^n,
    Y^n, Z^n) = (\mc, \mpr, \bfx, \bfy, \bfz)]} \nonumber\\*
  \quad &=& p_{\Mc,\Mpr} (\mc, \mpr)\, p_{X^n|\Mc,\Mpr} (\bfx|
  \mc,\mpr)\, \ch{W}^n (\bfy|\bfx)\, \ch{V}_{\bfs}
  (\bfz|\bfx), \label{eq:joint}
\end{IEEEeqnarray}
%\texttt{We are assuming here that $Y$ and $Z$ are conditionally
%  independent given the input and state.}
where
\begin{IEEEeqnarray}{rCl}
  p_{X^n|\Mc,\Mpr} (\bfx|\mc,\mpr) = \begin{cases}
    1 & \text{if $\bfx = f(\mc, \mpr)$} \\
    0 & \text{otherwise}
  \end{cases} \label{eq:encmap}
\end{IEEEeqnarray}
is determined by the encoder mapping;
\begin{IEEEeqnarray}{c}
  \ch{W}^n (\bfy | \bfx) = \prod_{i=1}^n \ch{W} (y_i | x_i);
\end{IEEEeqnarray}
and
\begin{IEEEeqnarray}{c}
  \ch{V}_{\bfs} (\bfz | \bfx) = \prod_{i=1}^n \ch{V}_{s_i} (z_i |
  x_i).
\end{IEEEeqnarray}

Fano's inequality yields that for any state sequence
$\bfs\in\set{S}^n$
\begin{subequations}
\label{block:Fanos}
\begin{IEEEeqnarray}{rCl}
  \log \card{\MMc} &\le& I(\Mc;Z^n) + n\eps_n  \label{eq:Fanos_a}\\
  \log \big( \card{\MMc} \card{\MMpr} \big) &\le& I(\Mc;Z^n) +
  I(\Mpr;Y^n) + n\eps_n \label{eq:fanosum} \label{eq:Fanos_b} \\
  \log \big( \card{\MMc} \card{\MMpr} \big) &\le& I(\Mc, \Mpr; Y^n) +
  n\eps_n, \label{eq:Fanos_c}
\end{IEEEeqnarray}
\end{subequations}
where $\eps_n$ approaches zero uniformly in $\bfs$ as $n$ tends to infinity.

% \texttt{Tricky. $\eps_{n}$ actually depends on $\bfs$. If we consider
%   randomized encoders we only know that---when averaged over the
%   random code selection---it tends to zero.}

For each $i\in[1:n]$ define
\begin{IEEEeqnarray}{rCl}
  U_i = (\Mc, Y_{i+1}^n, Z_1^{i-1}).
\end{IEEEeqnarray}
Let $Q$ be independent of $(\Mc, \Mpr, X^n, Y^n, Z^n)$ and uniformly
distributed over the integers $[1:n]$. Denote the joint PMF of $(Q,
\Mc, \Mpr, X^n, Y^n, Z^n)$ also by $\pmf{P}$, so
\begin{IEEEeqnarray}{rCl}
  \IEEEeqnarraymulticol{3}{l}{\pmf{P} [(Q, \Mc, \Mpr, X^n,
    Y^n, Z^n) = (q, \mc, \mpr, \bfx, \bfy, \bfz)]} \nonumber\\*
  \quad &=& \frac{1}{n}\, p_{\Mc,\Mpr} (\mc, \mpr)\, p_{X^n|\Mc,\Mpr}
  (\bfx| \mc,\mpr)\, \ch{W}^n (\bfy|\bfx)\, \ch{V}_{\bfs}
  (\bfz|\bfx), \label{eq:jointQ}
\end{IEEEeqnarray}
Define the chance variables
\begin{IEEEeqnarray}{rCl}
  \label{eq:auxrvs}
  U = (U_Q, Q),\quad X = X_Q,\quad Y = Y_Q,\quad Z = Z_Q,
\end{IEEEeqnarray}
and note that their joint distribution factorizes as in~\eqref{eq:factorizationForm}.

%\texttt{check!!!}

Continuing from~\eqref{block:Fanos}, we upper-bound
$\Rc = \frac{1}{n} \log \card{\MMc}$ and
$\Rc+\Rpr = \frac{1}{n} \log \big( \card{\MMc} \card{\MMpr} \big)$ by
the following calculations under the joint PMF $\ch{P}$
of~\eqref{eq:jointQ}. By~\eqref{eq:Fanos_a}
\begin{subequations}
\label{block:rcr}
\begin{IEEEeqnarray}{rCl}
  \Rc - \eps_n & \le & \frac{1}{n} I(\Mc; Z_1^{n}) \nonumber\\
  & = & \frac{1}{n} \sum_{i=1}^n I(\Mc; Z_i | Z_{1}^{i-1}) \label{step:rc1}\\
  & \le & \frac{1}{n} \sum_{i=1}^n I(\Mc, Y_{i+1}^n, Z_1^{i-1}; Z_i)
  \label{step:rc2}\\
  & = & I(U;Z|Q), \label{step:rc3}
\end{IEEEeqnarray}
\end{subequations}
where~\eqref{step:rc1} and~\eqref{step:rc2} follow from the chain rule
and the nonnegativity of mutual information, and~\eqref{step:rc3}
follows from the definitions in~\eqref{eq:auxrvs}.

By~\eqref{eq:Fanos_b},
%Similarly, the sum of the
%rates is upper-bounded by
\begin{subequations}
\label{block:sr}
\begin{IEEEeqnarray}{rCl}
  \Rc + \Rpr - \eps_n & \le & \frac{1}{n} \big( I(\Mc;Z_1^n) + I(\Mpr;
  Y_1^n | \Mc) \big) \label{step:sr1}\\
  & = & \frac{1}{n} \sum_{i=1}^n \big( I(\Mc;Z_i|Z_1^{i-1}) + I(\Mpr;
  Y_i |\Mc, Y_{i+1}^n) \big) \label{step:sr2}\\
  & \le & \frac{1}{n} \sum_{i=1}^n \big( I(\Mc;Z_i|Z_1^{i-1}) + I(X_i;
  Y_i |\Mc, Y_{i+1}^n) \big) \label{step:sr3}\\
  & \le & \frac{1}{n} \sum_{i=1}^n \big( I(\Mc, Z_1^{i-1};Z_i) +
  I(X_i, Z_1^{i-1}; Y_i |\Mc, Y_{i+1}^n) \big) \label{step:sr4}\\
  & = & \frac{1}{n} \sum_{i=1}^n \big( I(\Mc, Y_{i+1}^n,
  Z_1^{i-1};Z_i) - I(Y_{i+1}^n; Z_i|\Mc, Z_1^{i-1}) \nonumber\\
  && + I(Z_1^{i-1};Y_i|\Mc,Y_{i+1}^n) + I(X_i; Y_i |\Mc, Y_{i+1}^n,
  Z_1^{i-1}) \big) \label{step:sr5}\\
  & = & \frac{1}{n} \sum_{i=1}^n \big( I(\Mc, Y_{i+1}^n,
  Z_1^{i-1};Z_i) + I(X_i; Y_i |\Mc, Y_{i+1}^n, Z_1^{i-1}) \big) \label{step:sr6}\\
  & = & I(U;Z|Q) + I(X;Y|U,Q), \label{step:sr7}
\end{IEEEeqnarray}
\end{subequations}
where~\eqref{step:sr1} follows from~\eqref{eq:fanosum} (because $\Mc$
is independent of $\Mpr$); Equality~\eqref{step:sr2} follows from the
chain rule; Inequality~\eqref{step:sr3} follows because---conditional
on $(\Mc, Y_{i+1}^n)$---the chance variables
$\Mpr \markov X_i \markov Y_i$ form a Markov chain, i.e.,\footnote{In
  fact, we can replace the inequality with equality, because $X_{i}$
  is computable from $\Mc$ and $\Mpr$.}
\begin{equation*}
 \Mpr \markov (X_i,\Mc,Y_{i+1}^n) \markov Y_i; 
\end{equation*}
\eqref{step:sr4} and \eqref{step:sr5} follow from the chain rule and
the nonnegativity of mutual information; \eqref{step:sr6} follows from
Csisz\'ar's sum-identity; and \eqref{step:sr7} follows from
the definitions in~\eqref{eq:auxrvs}.

And finally, by~\eqref{eq:Fanos_c}
\begin{subequations}
\label{block:sr2}
\begin{IEEEeqnarray}{rCl}
  \Rc + \Rpr - \eps_n & \le & \frac{1}{n} I(\Mc, \Mpr; Y_1^n) \\
  & = & \frac{1}{n} I(X_1^n;Y_1^n) \label{step:eq1}\\
  & \le & \frac{1}{n} \sum_{i=1}^n I(X_i;Y_i) \label{step:iieq2}\\
  & = & I(X;Y|Q), \label{step:impfung12}
\end{IEEEeqnarray}
\end{subequations}
where~\eqref{step:eq1} holds because $X^n$ is a deterministic function
of $(\Mc, \Mpr)$; and~\eqref{step:iieq2} holds because the channel to
$Y$ is memoryless and without feedback.

Inequalities~\eqref{block:rcr}, \eqref{block:sr}, and
\eqref{block:sr2} hold for any state sequence $\bfs \in \set{S}^n$. We
next construct a specific state sequence and from it a joint PMF
$\tilde{p}_{U,X,Q}$. We will then show that the rates
$(\Rc - \eps_{n},\Rpr-\eps_{n})$ must satisfy~\eqref{block:eqvrregion}
when the latter is evaluated w.r.t. $\tilde{p}_{U,X,Q}$.
%for which $I(U;Z|Q)$ is
%less-than-or-equal to $\min_{s\in\set{S}} I(U;Z|Q)$ for a given
%$p_{U,X,Q}$ that is induced by the construction state sequence, the
%encoder, and the channel laws.
To construct the state sequence, we begin by expressing $I(U;Z|Q)$ as
a sum
\begin{IEEEeqnarray}{rCl}
  I(U;Z|Q) = \frac{1}{n} \sum_{i=1}^n I(U;Z|Q=i) = \frac{1}{n}
  \sum_{i=1}^n I(U_i;Z_i),
\end{IEEEeqnarray}
and consider each term separately in increasing order, starting with $i=1$.

The joint distribution $p_{U_1,X_1}$ is determined by the message
distribution $p_{\Mc, \Mpr}$, the encoder mapping~\eqref{eq:encmap},
and the channel law $\ch{W}$; it is uninfluenced by the state
sequence. We choose $s_1\in\set{S}$ so that
\begin{subequations}
\label{block:choicefirst}
\begin{IEEEeqnarray}{rCl}
  I(U_1,Z_1) = \min_{s\in\set{S}} I(U_1;Z),
\end{IEEEeqnarray}
where the mutual informations are computed w.r.t.\ the joint
distribution
\begin{IEEEeqnarray}{rCl}
  p_{U_1,X_1} (u_1,x_1)\, \ch{V}_{s_1} (z_1|x_1)\, \ch{V}_s (z|x_1).
\end{IEEEeqnarray}
\end{subequations}

Suppose now that we have chosen the first $i-1$ states
$s_{1}, \ldots, s_{i-1}$ for some $i \in [2:n]$. These states together
with the message distribution $p_{\Mc, \Mpr}$, the encoder
mapping~\eqref{eq:encmap}, and the channel law $\ch{W}$ fully specify
the joint distribution $p_{U_i,X_i}$ of $(U_{i},X_{i})$. 
%In general, for Term
%$i\in[2:n]$, the joint distribution $p_{U_i,X_i}$ is determined by the
%message distribuion $p_{\Mc, \Mpr}$, the encoder
%mapping~\eqref{eq:encmap}, the channel law $\ch{W}$, and our previous
%choices of the states $s_1, \ldots, s_{i-1}$.
We can then choose $s_i\in\set{S}$ so that
\begin{subequations}
\label{block:choicegeneral}
\begin{IEEEeqnarray}{rCl}
  I(U_i,Z_i) = \min_{s\in\set{S}} I(U_i;Z),
\end{IEEEeqnarray}
where the mutual informations are computed w.r.t.\ the joint
distribution
\begin{IEEEeqnarray}{rCl}
  p_{U_i,X_i} (u_i,x_i)\, \ch{V}_{s_i} (z_i|x_i)\, \ch{V}_s (z|x_i).
\end{IEEEeqnarray}
\end{subequations}
Once the entire state sequence $s_1, \ldots, s_n$ has been chosen, the
joint distributions $\{p_{U_i, X_i}\}_{i=1}^n$ are fully determined,
and we can define the joint distribution $\tilde{p}_{U,X,Q}$ as one
where $Q$ is uniform over $[1:n]$ and 
%and therefore so is $p_{U,X,Q}$, because
\begin{IEEEeqnarray}{rCl}
  \label{eq:puxq}
  \tilde{p}_{U,X|Q=i} = p_{U_i,X_i}, \qquad i \in [1:n].
\end{IEEEeqnarray}

We will next show that, under this joint distribution $\tilde{p}_{U,X,Q}$,
% $\{p_{U_i, X_i}\}_{i=1}^n$,
the rates $(\Rc - \eps_{n},\Rpr-\eps_{n})$ must
satisfy~\eqref{block:eqvrregion}.
Indeed, under this joint distribution, $I(U;Z|Q)$ can be upper-bounded as
\begin{subequations}
\label{block:minimization}
\begin{IEEEeqnarray}{rCl}
  I(U;Z|Q) & = & \frac{1}{n} \sum_{i=1}^n I(U;Z|Q=i) \\
  & = & \frac{1}{n} \sum_{i=1}^n I(U_i;Z_i) \\
  & = & \frac{1}{n} \sum_{i=1}^n \min_{s\in\set{S}}
  I(U_i;Z) \label{step:mindef} \\
  & \le & \min_{s\in\set{S}} \frac{1}{n} \sum_{i=1}^n
  I(U_i;Z) \label{step:minout} \\
  & = & \min_{s\in\set{S}} I(U;Z|Q), \label{eq:impfung117}
\end{IEEEeqnarray}
\end{subequations}
where~\eqref{step:mindef} follows from our choice of $s_i$
in~\eqref{block:choicefirst} and~\eqref{block:choicegeneral}, and
Inequality~\eqref{step:minout} holds because for the state
$s^{\star}\in\set{S}$ that minimizes the whole sum, the mutual
information $I(U_i;Z^{\star})$ w.r.t.\ the joint
$p_{U_i,X_i} (u_i,x_i)\, \ch{V}_{s^{\star}} (z^{\star}|x_i)$ is
greater or equal to $\min_{s\in\set{S}} I(U_i;Z)$ w.r.t.\ the joint
$p_{U_i,X_i} (u_i,x_i)\, \ch{V}_s (z|x_i)$ for all $i\in[1:n]$.

The upper bounds~\eqref{step:rc3}, \eqref{step:sr7},
\eqref{step:impfung12}, and~\eqref{eq:impfung117} together yield that,
under $\tilde{p}_{U,X,Q}$ of~\eqref{eq:puxq}, the rates of the coding
scheme are upper-bounded by
\begin{subequations}
\begin{IEEEeqnarray}{rCl}
  \Rc & \le & \min_{s\in\set{S}} I(U;Z|Q) + \eps_n \\
  \Rpr + \Rc & \le & I(X;Y|U,Q) + \min_{s\in\set{S}}
  I(U;Z|Q) + \eps_n \\
  \Rpr + \Rc & \le & I(X;Y|Q) + \eps_n,
\end{IEEEeqnarray}
\end{subequations}
where the mutual informations are computed w.r.t.\ joint PMF
\begin{IEEEeqnarray}{rCl}
  \tilde{p}_{U,X,Q} (u,x,q)\, \ch{W} (y|x)\, \ch{V}_s (z|x).
\end{IEEEeqnarray}
Having established that $(\Rc - \eps_{n}, \Rpr - \eps_{n})$ must
satisfy~\eqref{block:eqvrregion}, it now follows from the fact that
$\eps_{n}$ tend to zero and that $\CdetO$ is closed that $(\Rc,\Rpr)$
must be in $\CdetO$. And since the latter is equal to $\CdetI$
(Proposition~\ref{prop:R_is_CdetI}), the pair must also be in $\CdetI$.
\end{IEEEproof}

\section{Example}
Consider the binary symmetric semi-arbitrarily-varying broadcast
channel (BS-SAVBC), where the channel to~$Y$ is a $\BSC{p}$, i.e., a
binary symmetric channel (BSC) with crossover probability $p$,
% ---a channel that we denote $\BSC{p}$---
and the channel to~$Z$ is a BSC with a state-dependent crossover
probability between $\pmin$ and $\pmax$. The state alphabet $\set{S}$ is
the closed interval $[\pmin, \pmax]$, and we %sometimes
% refer to a generic element of $\set{S}$ by $p_s$. In fact, we
identify a state $s \in \set{S}$ with its corresponding crossover
probability $p_{s}$.  Thus, when the state is~$s$, %$p_s\in\set{S}$,
the channel from $X$ to $Z$ is a $\BSC{p_{s}}$.  We focus on the
case\footnote{When $p$ equals $1/2$ the capacity from $X$ to $Y$ is
  zero, and if we exclude this case, then---by possibily inverting
  $Y$---we can guarantee that $p$ be in $[0,1/2)$. Likewise, if the
  interval $[\pmin,\pmax]$ includes $1/2$, then the capacity of the
  AVC from $X$ to $Z$ is zero. And if this is excluded, then---again
  by possibly inverting $Z$---we can restrict ourselves to the case
  where this interval is a subset of $[0,1/2)$.
%  If $p=1/2$ or $\pmin \le 1/2 \le \pmax$, the
%  corresponding channel has zero capacity.
}
%assume without loss of generality\footnote{If $p=1/2$ or $\pmin \le 1/2 \le \pmax$, the
%  corresponding channel has zero capacity.}, that
\begin{IEEEeqnarray}{C}
  0 \le p < 1/2 \\
  0 \le \pmin \le \pmax < 1/2.
\end{IEEEeqnarray}

% More formally, the SAVBC is then given for every state $s\in\set{S}$
% by
% \begin{subequations}
% \begin{IEEEeqnarray}{rCl}
%   Y & = & X + N_{Y} \mod 2 \\
%   Z & = & X + N_{Z,s} \mod 2,
% \end{IEEEeqnarray}
% \end{subequations}
% where $X,Y,Z,N_{Y},N_{Z,s}$ are binary, and
% \begin{subequations}
% \begin{IEEEeqnarray}{rCl}
%   N_{Y} &\sim& \textnormal{Bernoulli} (p) \\
%   N_{Z,s} &\sim& \textnormal{Bernoulli} (p_s).
% \end{IEEEeqnarray}
% \end{subequations}

In this case the capacity of the DMC to $Y$ and of the AVC to $Z$ are
both positive (c.f.~\cite{CN88,Csi92}), and therefore (by
Remark~\ref{rem:nonemptyint} and Theorem~\ref{thm:rcapreg}) the
capacity region of the BS-SAVBC is $\CdetI$. Evaluating
\eqref{block:cap_noQ} for  the joint PMF $p_{U,X}$ under which
\begin{subequations}
\begin{IEEEeqnarray}{rCl}
  U &\sim& \textnormal{Bernoulli} (1/2) \\
  V &\sim& \textnormal{Bernoulli} (\alpha) \\
  X &=& U + V \mod 2,
\end{IEEEeqnarray}
\end{subequations}
proves that $\CdetI$ contains all the 
% the achievability of all
rate pairs $(\Rc, \Rpr)$ satisfying
\begin{subequations}
\label{block:bsbcreg1}
\begin{IEEEeqnarray}{rCl}
  \Rc & \le & \min_{s\in\set{S}} \big( 1 - \Hbin{\alpha \ast p_s}
  \big) \label{eq:bsbcreg11} \\
  \Rpr & \le & \Hbin{\alpha \ast p} - \Hbin{p} \\
  \Rc + \Rpr & \le & 1 - \Hbin{p}
\end{IEEEeqnarray}
\end{subequations}
for some $\alpha \in [0,1/2]$. Here $\Hbin{\cdot}$ denotes the binary
entropy function, and
$\alpha \ast \delta = \alpha(1-\delta) + (1-\alpha)\delta$.

For a fixed $\alpha\in [0,1/2]$ the mapping
$\delta \mapsto \alpha \ast \delta$ is nondecreasing on
$(0 < \delta < 1/2)$, and so is $\Hbin{\cdot}$.  Consequently, the
minimum on the RHS of~\eqref{eq:bsbcreg11} is achieved when $p_s$
equals $\pmax$, and~\eqref{block:bsbcreg1} simplifies to
\begin{subequations}
\label{block:bsbcreg2}
\begin{IEEEeqnarray}{rCl}
  \Rc & \le & 1 - \Hbin{\alpha \ast \pmax} \label{eq:bsbcreg21} \\
  \Rpr & \le & \Hbin{\alpha \ast p} - \Hbin{p} \label{eq:bsbcreg22} \\
  \Rc + \Rpr & \le & 1 - \Hbin{p}. \label{eq:bsbcreg23}
\end{IEEEeqnarray}
\end{subequations}

\begin{figure}
\makebox[\textwidth][c]{\includegraphics[scale=0.5]{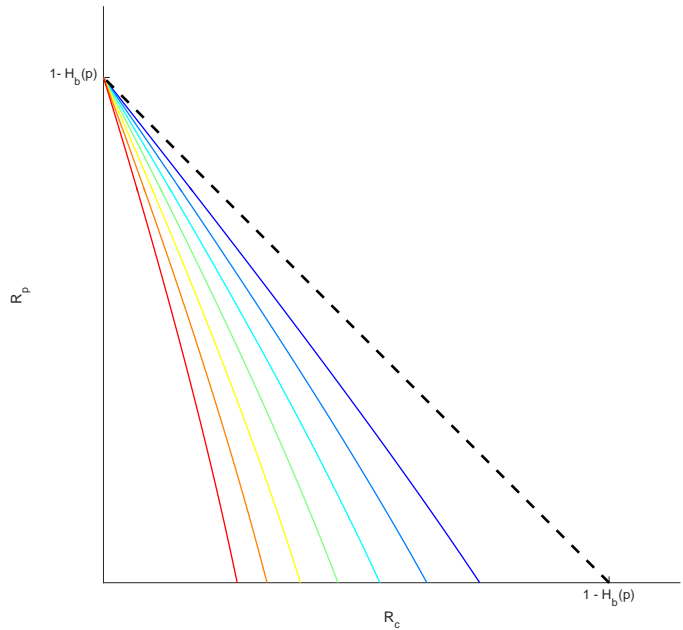}}
\caption{The boundary of the capacity region of the binary symmetric semi-arbitrarily
  varying broadcast channel for various values of $\pmax > p$. The
  capacity region shrinks (and eventually has an empty interior)
  as $\pmax$ increases to $1/2$. If $\pmax \le p$, the capacity region
  is the triangle defined by the sum-rate constraint $\Rc + \Rpr \le
  1-\Hbin{p}$.
  \label{fig:SAVBSBC}
}
\end{figure}

We next show that $\CdetI$ contains no other rate pairs, and that it
thus equals the union over all $\alpha \in [0,1/2]$ of the polytopes
defined by~\eqref{block:bsbcreg2}. This region is depicted
in~Figure~\ref{fig:SAVBSBC}.
We do so by fixing the state to be $\pmax$ throughout the block and by
then showing that every achievable rate pair $(\Rc, \Rpr)$ must
satisfy~\eqref{block:bsbcreg2} for some $\alpha \in [0,1/2]$. To this
end, we distinguish between two cases, depending on whether or
not $p$ exceeds $\pmax$.

But first we note that, by the above monotonicity argument, the
relation between $p$ and $\pmax$ translates to the relation between
$\Hbin{\alpha \ast p}$ and $\Hbin{\alpha \ast \pmax}$ as follows:
\begin{subequations}
\begin{IEEEeqnarray}{rCl}
  \Big( p \le \pmax \Big) &\iff& \Big( \Hbin{\alpha \ast p} \le
  \Hbin{\alpha \ast \pmax} \Big) \label{eq:caseI}\\
  \Big( p > \pmax \Big) &\iff& \Big( \Hbin{\alpha \ast p} >
  \Hbin{\alpha \ast \pmax} \Big). \label{eq:caseII}
\end{IEEEeqnarray}
\end{subequations}

%We are now ready to prove that $\CdetI$ contains only rate pairs that
%satisfy~\eqref{block:bsbcreg2} for some $\alpha \in [0,1/2]$. 
% By separately considering two cases, depending on whether or not $p$
% exceeds $\pmax$, we will show that the inner bound defined
% by~\eqref{block:bsbcreg2} is tight and therefore characterizes the
% capacity region, which is plotted in Figure~\ref{fig:SAVBSBC}.

\noindent
\emph{Case I: $p \le \pmax$.}

\nopagebreak
\noindent
In this case fixing the state at $\pmax$ results in a stochastically
degraded binary-symmetric broadcast channel (BS-BC), where $Z$ is a
stochastically degraded version of $Y$. Since Receiver~$Y$ recovers
$(\Mc, \Mpr)$, and Receiver~$Z$ recovers~$\Mc$, any achievable rate
pair $(\Rc, \Rpr)$ must be in the private-message capacity
region of the above BS-BC. The latter is given by the set of rate
pairs $(\Rc, \Rpr)$ that satisfy
\begin{subequations}
\label{block:degradedBC}
\begin{IEEEeqnarray}{rCl}
  \Rpr &\le& I(X;Y|U) \\
  \Rc &\le& I(U;Z)
\end{IEEEeqnarray}
\end{subequations}
for some PMF $p_{U,X}$~\cite[Theorem 5.2]{ElGamalKim}. For the
stochastically degraded BS-BC with the stronger receiver $Y$ observing
the BSC($p$) and the degraded receiver $Z$ observing the BSC($p_s$),
the capacity region~\eqref{block:degradedBC} simplifies to the set of
rate pairs $(\Rc, \Rpr)$ that satisfy
\begin{subequations}
\label{block:bsbcregO}
\begin{IEEEeqnarray}{rCl}
  \Rpr & \le & \Hbin{\alpha \ast p} - \Hbin{p} \\
  \Rc & \le & 1 - \Hbin{\alpha \ast \pmax}
\end{IEEEeqnarray}
\end{subequations}
for some $\alpha\in [0, 1/2]$~\cite[Section 5.4.2]{ElGamalKim}. Since
these inequalities coincide with~\eqref{eq:bsbcreg21}
and~\eqref{eq:bsbcreg22}, it follows that to every rate pair
$(\Rc, \Rpr) \in \CdetI$
%
%It now follows from~\eqref{block:bsbcregO} that to any achievable rate
%pair $(\Rc, \Rpr)$
there corresponds some $\alpha\in [0, 1/2]$ for
which~\eqref{eq:bsbcreg21} and~\eqref{eq:bsbcreg22} are satisfied. The
sum-rate constraint~\eqref{eq:bsbcreg23} is satisfied automatically
because, in the case at hand, \eqref{eq:bsbcreg21}
and~\eqref{eq:bsbcreg22} imply~\eqref{eq:bsbcreg23}. Indeed,
adding~\eqref{eq:bsbcreg21} and \eqref{eq:bsbcreg22} yields
\begin{IEEEeqnarray}{rCl}
  \Rc + \Rpr & \le & 1 - \Hbin{\alpha \ast \pmax} + \Hbin{\alpha \ast p} -
  \Hbin{p} \\
  & \le & 1 - \Hbin{p},
\end{IEEEeqnarray}
where the second inequality follows from~\eqref{eq:caseI}.

% %\begin{IEEEeqnarray}{rCl}
% %  \Rc + \Rpr \le 1 - \Hbin{\alpha \ast \pmax} + \Hbin{\alpha \ast p} -
% %  \Hbin{p} \le 1 - \Hbin{p}.
% %\end{IEEEeqnarray}
% The rate pairs satisfying~\eqref{block:bsbcreg2} are thus those that satisfy
% \begin{subequations}
% \label{block:bsbcregO}
% \begin{IEEEeqnarray}{rCl}
%   \Rc & \le & 1 - \Hbin{\alpha \ast \pmax} \\
%   \Rpr & \le & \Hbin{\alpha \ast p} - \Hbin{p}
% \end{IEEEeqnarray}
% \end{subequations}
% for some $\alpha\in [0, 1/2]$. We next argue, that this inner bound
% for the capacity region is tight.

% We first note that the region defined by~\eqref{block:bsbcregO} is the
% \emph{private-message} capacity region of the
% % deterministic (non-varying)
% binary symmetric broadcast channel (BSBC) with crossover probabilities
% $p$ and $\pmax$~\cite[Section 5.4.2]{ElGamalKim}. And since under the
% case at hand $p \leq \pmax$, this broadcast channel is stochastically
% degraded with $Y$ being the better receiver that decodes the
% rate-$\Rpr$ stream, and $Z$ being the degraded receiver that decodes
% the rate-$\Rc$ stream.

% Also note that in our setting with a common message and only one
% private message, we can view the common message as the private message
% for the robust receiver (that is also decoded by the ordinary
% receiver). Consequently, since the jammer can select the state
% sequence to be constat equal to $\pmax$, the rate pair $(\Rc, \Rpr)$
% must lie in the capacity region of the BSBC with $Y$ observing the
% BSC($p$) and $Z$ observing the BSC($p_s$) that is characterized
% by~\eqref{block:bsbcregO}.

\noindent
\emph{Case II: $p > \pmax$.}

\nopagebreak
\noindent
In this case fixing the state at $\pmax$ again results in a
stochastically degraded BS-BC, but in reverse order: now
%
%In this case when the state is fixed at $\pmax$ we again obtain a
%stochastically degraded BS-BC, but this time with
$Y$ is a degraded version of $Z$. To show that any achievable rate
pair $(\Rc, \Rpr)$ must satisfy~\eqref{block:bsbcreg2}, we first note
that---since it is now the weaker receiver, namely Receiver~$Y$, that
must recover both $\Mc$ and $\Mpr$---the sum-rate $\Rc + \Rpr$ must
not exceed the Shannon capacity of the BSC($p$) from $X$ to $Y$
%\begin{subequations}
  \begin{IEEEeqnarray}{rCl}
  \Rc + \Rpr & \le & 1 - \Hbin{p}. \label{eq:bsbcreg23_amos}
\end{IEEEeqnarray}
% \end{subequations}
Every rate pair in $\CdetI$ must thus
satisfy~\eqref{eq:bsbcreg23_amos}.

We next show that, to every rate pair $(\Rc, \Rpr)$
satisfying~\eqref{eq:bsbcreg23_amos}, there corresponds some
$\alpha \in [0,1/2]$ for which~\eqref{block:bsbcreg2} hold.  To see
why note that, for the case at hand, for every $\alpha \in [0,1/2]$
the pair
\begin{subequations}
  \label{block:maspik}
\begin{IEEEeqnarray}{rCl}
  \Rc & = & 1 - \Hbin{\alpha \ast p} \\
  \Rpr & = & \Hbin{\alpha \ast p} - \Hbin{p}
\end{IEEEeqnarray}
\end{subequations}
satisfies \eqref{block:bsbcreg2} (because, by~\eqref{eq:caseII},
$1 - \Hbin{\alpha \ast p}$ cannot exceed
$1 - \Hbin{\alpha \ast \pmax}$ and~\eqref{eq:bsbcreg21} must therefore
hold). As we vary $\alpha$ from $0$ to $1/2$, the rate
pair~\eqref{block:maspik} traces the line $\Rc+\Rpr = 1 - \Hbin{p}$.

\bibliographystyle{IEEEtran}
\bibliography{./all}

\end{document}